\documentclass[10pt,letterpaper]{article}
\usepackage{opex3}
\usepackage{array}
\usepackage{subfig}
\usepackage{bm}
\usepackage{amsfonts}
\usepackage{amsmath}
\usepackage{epsfig}

\def\bx{{\bar x}}

\def\bE{\mathbb{E}}
\def\RR{\mathbb{R}}

\def\cH{{\cal H}}

\def\erfc{{\rm erfc\, }}

\def\be{\begin{equation}}
\def\ee{\end{equation}}
\def\ba{\begin{align}}
\def\defeq{\buildrel \rm def \over =}
\newcommand{\argmax}[1]{\underset{#1}{\operatorname{arg}\operatorname{max}}\;}

\newcommand{\norm}[1]{\left\lVert#1\right\rVert_2}

\def\MPE{P_e^{(min)}}
\def\cR{{\cal R}}
\def\cRm{{{\cal R}_m}}
\def\cRmp{{{\cal R}_{m^\prime}}}

\def\tm{{\tilde m}}
\begin{document}
\title{\bf Asymptotics of Bayesian Error Probability and 2D Pair Superresolution}
\author{S.~Prasad}%
\address{Department of Physics
and Astronomy, University of New Mexico, Albuquerque, New Mexico 87131}
\email{sprasad@unm.edu}

\begin{abstract}
This paper employs a recently developed asymptotic Bayesian multi-hypothesis testing (MHT) error
analysis \cite{SP14} to treat the problem of superresolution imaging of a pair of closely spaced, equally bright
point sources. The analysis exploits the notion of the minimum probability of error (MPE) in discriminating between
two competing equi-probable hypotheses, a single point source of a certain brightness at the origin vs. a pair of point sources, each
of half the brightness of the single source and located symmetrically about the origin, 
as the distance between the source pair is changed. For a Gaussian point-spread function (PSF), the 
analysis makes predictions on the scaling of the minimum source strength, expressed in units of photon number,
required to disambiguate the pair as a function of their separation, in both the signal-dominated and background-dominated
regimes. Certain logarithmic corrections to the quartic scaling of the minimum source strength with respect to
the degree of superresolution characterize the signal-dominated regime, while the scaling is purely quadratic in the 
background-dominated regime.
For the Gaussian PSF, general results for arbitrary strengths of the signal, background, and sensor noise levels 
are also presented. 
\end{abstract}



\section{Introduction}
Recent ground-breaking advances in fluorescent biomarker microscopy \cite{HBZ09} have enabled the localization and tracking of
single molecules at scales of mere tens of nanometers, scales that are 10-20x finer than those achieved by the more traditional 
diffraction limited microscopy. As the need for rapid localization and tracking has grown, the question of disambiguation
of multiple point sources in close vicinity of one another has also become important. At the most elementary level,
one must characterize well the minimum SNR requirements on finely resolving a single pair of equally bright
point sources with a given sub-diffractive separation under varying levels of background, sensor, and signal noise. 
Such a characterization is critical to the design and construction of ever more capable microscopes that must perform these
tasks with exquisite spatio-temporal resolution in our effort to understand biological processes at the cellular level.   

We recently analyzed \cite{SP14} the minimum probability of error (MPE) in the localization of a single molecule to sub-diffractive scales.
Our statistical approach employed the Bayesian multi-hypothesis testing (MHT) protocol that for the problem of transverse (2D) localization 
assigns {\it a priori} equal probability, $1/M_\perp^2$, that the molecule is located in any one of $M_\perp\times M_\perp$ possible 
equal-area square subcells into which a chosen square base-resolution cell may be subdivided. The image data, essentially the
point-spread function (PSF) of the imaging system corrupted by a variety of noise sources, including background, sensor, and photon
noise, add information about the source location, helping thus to reduce the MPE from $1-1/M_\perp^2$, 
that for the equi-probable prior alone, to a value that can be bounded below 
by means of the maximum {\it a posteriori} probability (MAP) criterion for choosing the 
decision regions. If the data are of a sufficiently high quality, then the MPE can dip below a certain threshold value, say
5\% corresponding to a statistical confidence limit (CL) of 95\%, for localizing the molecule with $M_\perp$ times higher 
linear precision along each transverse dimension. This 2D localization error analysis was extended to include the axial, or depth, dimension as well. 
A complete analysis, including its asymptotics, of the minimum signal strengths needed for super-localizing a point source to ever 
smaller spatial scales in both transverse and axial dimensions under varying background and sensor noise conditions
was presented in Ref.~\cite{SP14}. 
    
The present paper specializes our MHT-based MPE analysis to the related problem of superresolution imaging of 
multiple point sources that are closely separated and thus subject to being confused as a single point source
of the same intensity as the combined intensity of the individual sources. 
This is particularly true if the spacing is well within the Abbe diffraction limit, $0.61\lambda/NA$,
of a microscope operating with numerical aperture $NA$ and imaging wavelength $\lambda$.
However, as a number of recent experimental results have shown \cite{PDML10,KPS14,GM14}, 
the diffraction limit furnishes
a nominal benchmark at best, and improvements well beyond it are eminently possible
based on diffraction-limited far-field image data. 

Here we ask and answer, in the Bayesian MPE framework, the elementary question: 
What source brightness is needed to reduce the 
MPE below a given threshold for discriminating a single pair of closely spaced, equally bright point sources of 
a given separation from a single point source of the same brightness located at
the midpoint of the line joining the pair? This defines a simpler problem of {\it binary} hypothesis
testing within the Bayesian detection framework, avoiding 
a more involved problem in which the matter of such binary {\it discrimination} is coupled with
the additional requirement of {\it estimating} the spacing between the point-source pair.
The minimum signal strengths needed for this single source vs. pair source discrimination, as we shall see,
are rather modest and consistent with the growing experimental evidence for high-precision 
optical superresolution (OSR) of point sources.

Similar questions of source-pair OSR in the framework of Bayesian inference
have been asked previously by others \cite{Harris64,Helstrom69,vdB02,SM04}. These
analyses are more limited, however, because they either employ simpifying assumptions such as one-dimensional
PDFs or do not account for non-Gaussian sources of noise and consider
the full range of noise sources from additive sensor noise to signal and background-based shot
noise comprehensively. Considerations of the minimum requirements for superresolution
of closely spaced point sources have recently gained prominence in view
of highly developed experimental technqiues for superresolution fluorescence microscopy \cite{HBZ09} for localizing single 
biolabel fluorophors that are now quite the rage in live-cell biological and biomedical imaging.

The limits on the spatial resolution of a closely spaced pair of point sources have been examined previously \cite{CRWO09}
in the context of biomolecular microscopy
by means of a non-Bayesian estimation-theoretic analysis involving the Cram\'er-Rao lower bound (CRB) on an unbiased estimation
of their mutual spacing. This analysis, while exhaustive in terms of the orientation
and the location of the midpoint of the separation vector near the in-focus plane of 
a conventional clear-circular-aperture imager, was largely numerical, providing only limited insights 
into the minimum requirements, e.g., on the combined source strength, to resolve a specific 
subdiffractive separation in the presence of varying amounts of background and sensor noise levels.  
Even more importantly, this analysis suffers from the same weakness as any other CRB-based analysis in
requiring certain regularity conditions for its validity, including a non-vanishing local first-order derivative of the 
statistical distribution of the data with respect to (w.r.t.) the parameter being estimated.
This specific condition breaks down when the sources are located along the optical axis and their separation vector is centered
at the in-focus plane where the imager PSF has a vanishing first-order derivative w.r.t. 
the axial coordinate and the CRB in fact diverges \cite{RWO05} as a result. 

The general OSR problem subsumes a large collection of well defined problems, a fact that merits
some discussion here. The many claims and counter-claims of the possibility of OSR debated over the
past several decades have as much to do with the complexity
of proper analysis as with the nature of the specific OSR problem being considered. 
At the risk of over-simplification, we sharpen this debate with the mention of two specific OSR problems here.
For one, the prohibitively difficult prospects of bandwidth extension beyond the diffraction limited cut-off
by means of data inversion performed by imposing physical constraints \cite{Kosarev90,RH68,BdM96,BL03,MT06,PL09} 
constitute a very different problem from 
that of superresolving point sources. The degree of bandwidth extension is typically logarithmic in the signal-to-noise ratio (SNR),
a fact that is also well understood from Shannon's channel capacity theorem \cite{Kosarev90}.
But the knowledge that all sources in a scene are point-like, rather than
being extended with arbitrary variations of brightness across them, is sufficiently information-bearing
that superresolving them to well within the diffraction limit requires, as we hope to prove, 
rather more modest source-signal strengths. Independent confirmations of this claim are provided by a number of studies 
involving actual computational 
algorithms, including those presented in \cite{Gerchberg74,Fried95,MCS98,SPM02,PH08}. These algorithms make use,
in essential ways, of the point-like extension of sources being reconstructed from noisy, filtered image data.

It is also worth noting that Lucy's early analysis \cite{Lucy92} of
the point-source OSR problem in which he derived a prohibitive eighth-power scaling of the minimum source strength 
w.r.t. the pair separation was based on his asking a different question: What source strength is needed
to disambiguate this pair from a single {\it extended}, spatially symmetric source
of the same source strength and the same second-order spatial moment of the intensity distribution?  
Here the alternative (``null") hypothesis involves an extended source of the same 
root-mean-squared (RMS) width as the separated source-pair in the main hypothesis.
It is this additional constraint of equal RMS width placed on the two hypotheses that
renders pair OSR so much harder to achieve in his analysis.    

We begin this paper with a brief review of our earlier analysis \cite{SP14} 
of the MPE as applied to the problem of super-localizing a single point source in three dimensions.
The specialization of this multi-hypothesis testing (MHT) analysis and its asymptotic
considerations to the present problem of pair OSR in the single transverse plane of best focus
appear in Sec. 3, where we further specialize the problem to the case of a Gaussian model PSF. 
This restriction to the Gaussian PSF shape, rather than the Bessel-function-based Airy-disk PSF, is not
essential but does enable a more transparent and complete analytical treatment. 
The biomolecular localization microscopy literature \cite{TLW02,HBZ09} on the specific form 
of the in-focus PSF is divided, since the presence of any residual aberrations in the optics and
the dipolar emission pattern \cite{ETS06} from a flourophore typically
predispose a single-molecule image to deviate significantly from the Airy form, so the more
generic bell-shaped form is as good, typically, as any other to assume. Our considerations
may, however, be adapted to a completely general image form by means of a Taylor
expansion to the second order along the lines of References \cite{SM04,RWO06}.
For the Gaussian PSF, we derive detailed expressions for the MPE and also their asymptotic
form for signal and background dominated regimes, including an important logarithmic 
correction to the quartic scaling law for the minimum photon strength needed for
a sought pair OSR enhancement factor in the strong-signal regime.   
In Sec. 4, we present and discuss the results of a numerical evaluation of 
our analytical expressions, involving certain numerically ill-behaved 
logarithmic integrals derived in Sec. 3, for arbitrary relative strengths
of the signal and background levels. We conclude the paper in Sec. 5 with a brief summary
and outlook of the work. 

\section{A Brief Review of our Previous Asymptotic MHT Analysis}
 
Let the conditional statistics of the data $X$ be specified by a probability density function (PDF), $P(x|m)$, 
conditioned on the validity of a specific hypothesis $\cH_m$, labeled by an integer $m$, $m=1,\ldots,M$. 
In the Bayesian framework, the posterior distribution, $P(m|x)$, quantifies the information carried
by the data about the likelihood of the hypothesis $\cH_m$ having given rise to the observed data.    
The maximum {\it a posteriori} (MAP) estimator provides one
fundamental metric of minimum error, namely the MPE, in correctly identifying
the operative hypothesis under all possible observations, 
\be
\label{e0}
\MPE=1-\bE\left[P(\hat m_{MAP}\mid X)\right],
\ee
where $\hat m$ is the MAP estimator,
\be
\label{e1}
\hat m_{MAP}(X)=\argmax{m=1,\ldots,M}P(m\mid X).
\ee
Expression (\ref{e0}) may be reformulated by means of the Bayes rule as a double sum of data integrals,
\be
\label{e2}
\MPE=\sum_{m=1}^M p_m\sum_{m^\prime\neq m}\int_\cRmp dx \, P(x\mid m)
\ee
over the various decision regions $\cRmp$ chosen according to the MAP protocol.

When many observations are involved, as, e.g., for the typical image dataset consisting of $N$ pixels with
$N>>1$, $\MPE$ may be evaluated approximately under conditions of moderate to high SNR 
by replacing the inner sum over $m^\prime$ in Eq.~(\ref{e2}) by a single
term $\tilde m$ that labels the decision region ``closest" to $\cRm$ in the following sense:
\be
\label{e3}
\tilde m = \argmax{m^\prime\neq m} \max_{x\in \cRmp}\left\{P(x\mid m)\right\}.
\ee
The MPE is thus accurately approximated by the asymptotic expression
\be
\label{e4}
\MPE=\sum_{m=1}^M p_m\int_{{\cal R}_{\tilde m}} dx \, P(x\mid m)
\ee
at sufficiently high values of the SNR.

For the case of image data acquired under combined signal photon-number fluctuations, background fluctuations,
and sensor read-out noise, the following pseudo-Gaussian conditional data PDF 
accurately describes the statistics of data at least at large photon numbers: 
\be
\label{e5}
P(x\mid m) ={1\over (2\pi)^{N/2}{\rm det\,}^{1/2}(\Sigma_m)}\exp[-(1/2)(x^T-x_m^T)\Sigma_m^{-1}(x-x_m)].
\ee
where, under the condition of statistically independent data pixels, the data covariance matrix is a diagonal matrix 
of the form\footnote{ 
We use here a shorthand notation, diag$(v)$, for specifying a diagonal matrix whose  
diagonal elements are the elements of $v$ taken in order. Similarly, diag$(u/v)$, denotes 
the diagonal matrix of elements that are ratios of the corresponding elements of the vectors $u$ and $v$. In Matlab, this would
be the element-wise quotient, $u./v$, of the two vectors of which the diagonal matrix is formed.} 
\be
\label{e6}
\Sigma_m ={\rm diag}(\sigma^2+b+x_m),
\ee
where $\sigma_2$, $b$, and $x_m\in \RR^N$ denote, respectively, the variance of sensor read-out noise, the mean
background count, assumed spatially uniform, and the mean signal vector, given the hypothesis $m$.

For this problem, we derived in Ref. \cite{SP14} the following expression valid in the asymptotic limit
of high photon numbers, many pixels, and hypotheses that are hard to discriminate from one another because
the corresponding mean signal vectors that separate their statistics are very similar:
\be
\label{e7}
\MPE={1\over 2}\sum_m p_m\, {\rm erfc}\, (\norm{U_m}/\sqrt{2}), 
\ee
where $\norm{U_m}$ takes the asymptotic form
\be
\label{e8}
\norm{U_m}={1\over 2}{\sum_{i=1}^N{(\sigma^2+b+x_{mi})^{1/2}
\over (\sigma^2+b+\bx_{mi})^2}(\delta x_{\tm mi})^2
\over 
\left[\sum_{i=1}^N{1 \over (\sigma^2+b+\bx_{mi})^2}(\delta x_{\tm mi})^2\right]^{1/2}}
\ee
in terms of the components of the vector separation, $\delta x_{\tm m}$, between the mean data vectors
for the nearest-neighbor decision-region pairs $\cRm$ and $\cR_\tm$, as defined by relation (\ref{e3}). 
The arithmetic mean, $\bx_m$, of the mean signal vectors in the two decision regions that 
occurs in this expression may be replaced by either mean signal vector, say $x_m$, as we do presently,
without incurring significant error in the asymptotic limit as defined above.

\section{Bayesian MPE Analysis for the Point-Source-Pair Superresolution Problem}

As we have indicated earlier, the problem of discriminating a pair of closely spaced point sources from 
a single point source is a binary hypothesis testing (BHT) problem in which the sum 
(\ref{e8}) is limited to two terms only, $m=1$ and $m=2$ corresponding to the 
cases of a single point source and a pair of point sources, respectively.  
Since the passage from the double sum (\ref{e2}) to the single sum (\ref{e4})
is exact for the BHT problem, any error involved in the expression (\ref{e7})
is only due to any asymptotic analysis of the full MPE contribution from each decision region.
In the asymptotic limit and for only two terms in the sum (\ref{e7}), the two norms, $\norm{U_1}$
and $\norm{U_2}$, as defined by relation (\ref{e8}) are essentially the same, namely
\be
\label{e9}
\norm{U_0}\defeq {1\over 2}{Q\over R^{1/2}},
\ee
where $Q$ and $R$ are defined as the sums
\ba
\label{e9a}
Q\defeq &\sum_{i=1}^N{1
\over (\sigma^2+b+x_{1i})^{3/2}}(x_{2i}-x_{1i})^2; \nonumber\\
R\defeq & \sum_{i=1}^N{1 \over (\sigma^2+b+x_{1i})^2}(x_{2i}-x_{1i})^2.
\end{align}
Since the priors add to 1, $p_1+p_2=1$, the MPE (\ref{e7}) thus reduces to the simple form
\be
\label{e10}
\MPE={1\over 2}{\rm erfc}\, (\norm{U_0}/\sqrt{2}). 
\ee
For the 2D imaging problem at hand, the single sum over the elements $i$ in each of our above expressions expands into a double 
sum over the pixel indices $i$ and $j$.

In the following analysis we shall restrict our attention to a circular Gaussian-shaped PSF that is
azimuthally invariant in the $\xi\eta$ image plane,
\be
\label{e11}
H(\xi,\eta) = {1\over 2\pi w^2} \exp\left[-{(\xi^2+\eta^2)\over 2w^2}\right],
\ee
which is normalized to have unit volume, $\int\int d\xi\, d\eta \, H(\xi,\eta) = 1$,
as appropriate for the PSF of a lossless imager. The continuous PSF (\ref{e11})
may be approximated by its discrete form, taking the following value 
on the $(i,j)$th pixel, centered  the point $(\xi_{ij},\eta_{ij})$:
\be
\label{e12}
H_{ij} = {1\over 2\pi w^2} \exp\left[-{(\xi_{ij}^2+\eta_{ij}^2)\over 2w^2}\right] \Delta A,
\ee
in which $\Delta A$ is the area of each of the many square pixels over which the PSF is assumed
to be distributed. In the discrete form (\ref{e12}), the PSF is normalized approximately
as
\be
\label{e13}
\sum_{i,j} H_{ij} =1,
\ee 
a relation that becomes exact in the limit $\Delta A\to0$. 

The two hypotheses in the present problem are characterized by the mean signals, $x_1$ and $x_2$,
that are given in terms of the discrete PSF (\ref{e12}) as 
\ba
\label{e14}
x_{1ij}=&K\, H_{1ij} = {K\over 2\pi w^2} \exp\left[-{\xi_{ij}^2+\eta_{ij}^2\over 2w^2}\right] \Delta A; \nonumber\\
x_{2ij}=& K\, H_{2ij}\nonumber\\ 
=&{K\over 4\pi w^2} \left\{\exp\left[-{(\xi_{ij}-d/2)^2+\eta_{ij}^2\over 2w^2}\right]
 +\exp\left[-{(\xi_{ij}+d/2)^2+\eta_{ij}^2\over 2w^2}\right]\right\}\Delta A,
\end{align}
in which $K$ is the source strength in units of photon number, $d$ is the spacing between the point sources, 
assumed to be situated on the $\xi$ axis, and 
the two data mean vectors have expanded to become matrices supported on the pixel array for the 2D imaging problem. 
The quantum efficiency (QE) is assumed to be 1 here, but an imperfect QE is easily included
by multiplying both $K$ and $b$ by it in all of the expressions.
Figure 1 shows examples of the image (\ref{e14}) for the point-source pair for three
different values of the ratio $d/w$.
\begin{figure}
\centering
\subfloat[]
{\includegraphics[width=2in]{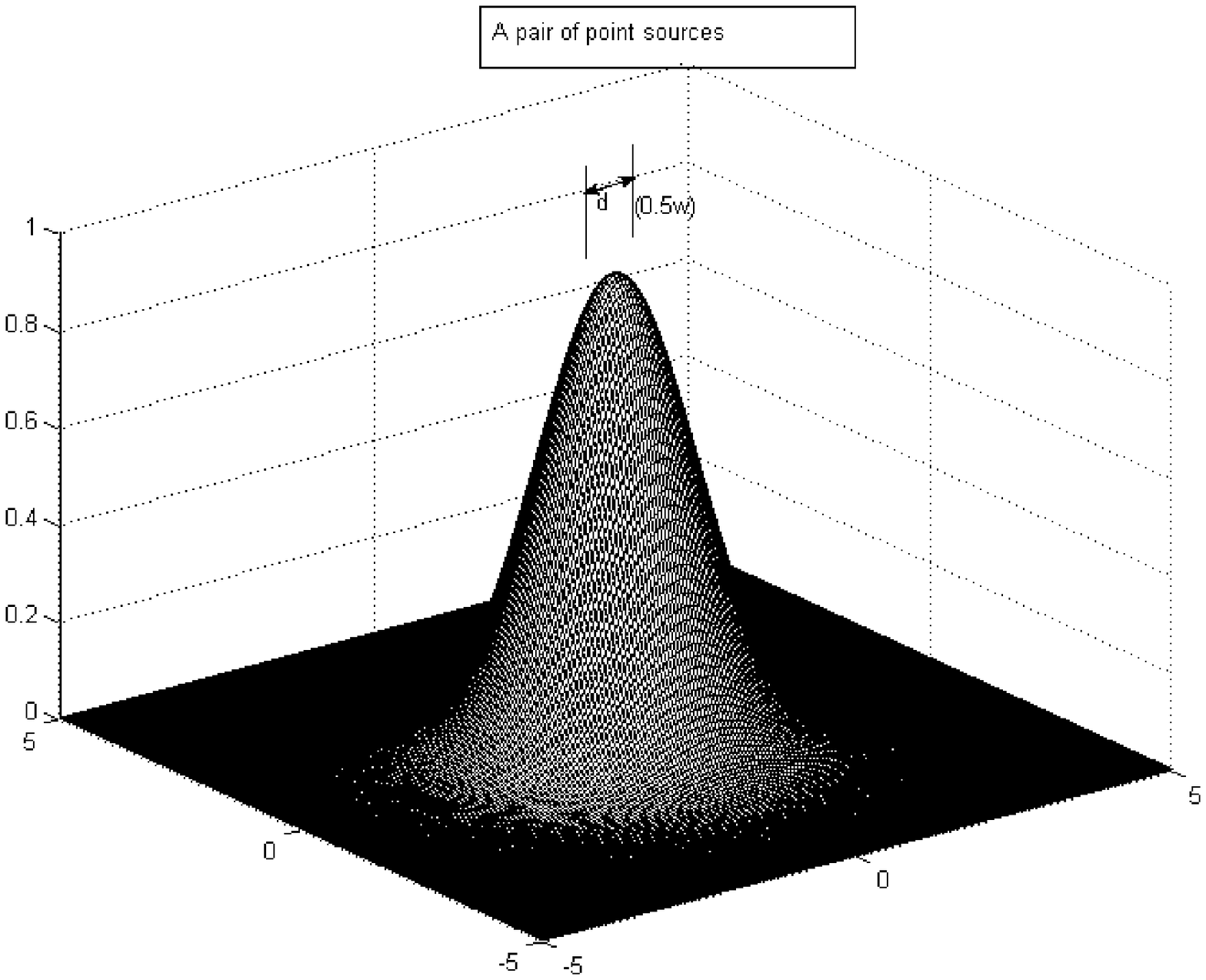}}
\subfloat[]
{\includegraphics[width=2in]{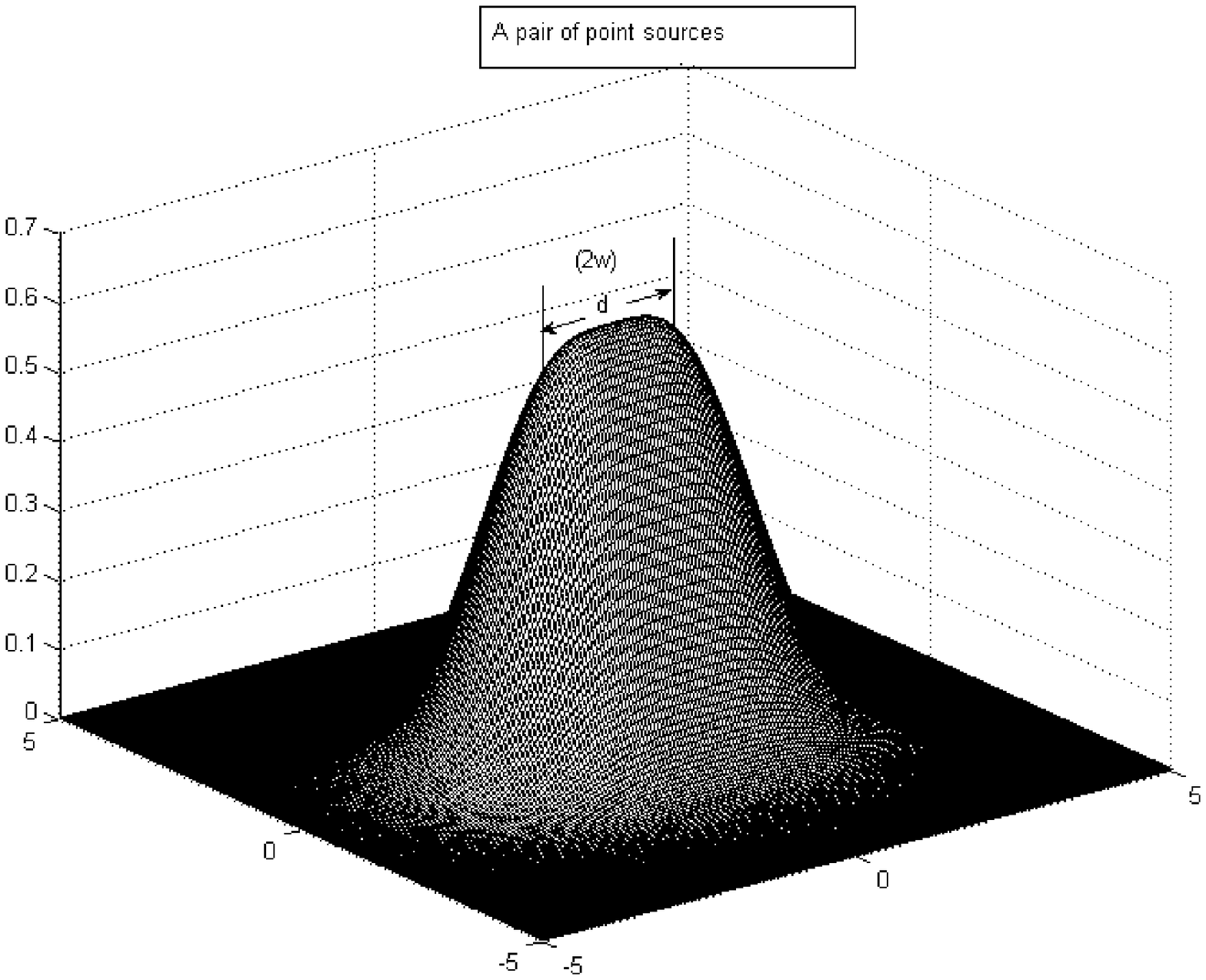}}
\subfloat[]
{\includegraphics[width=2in]{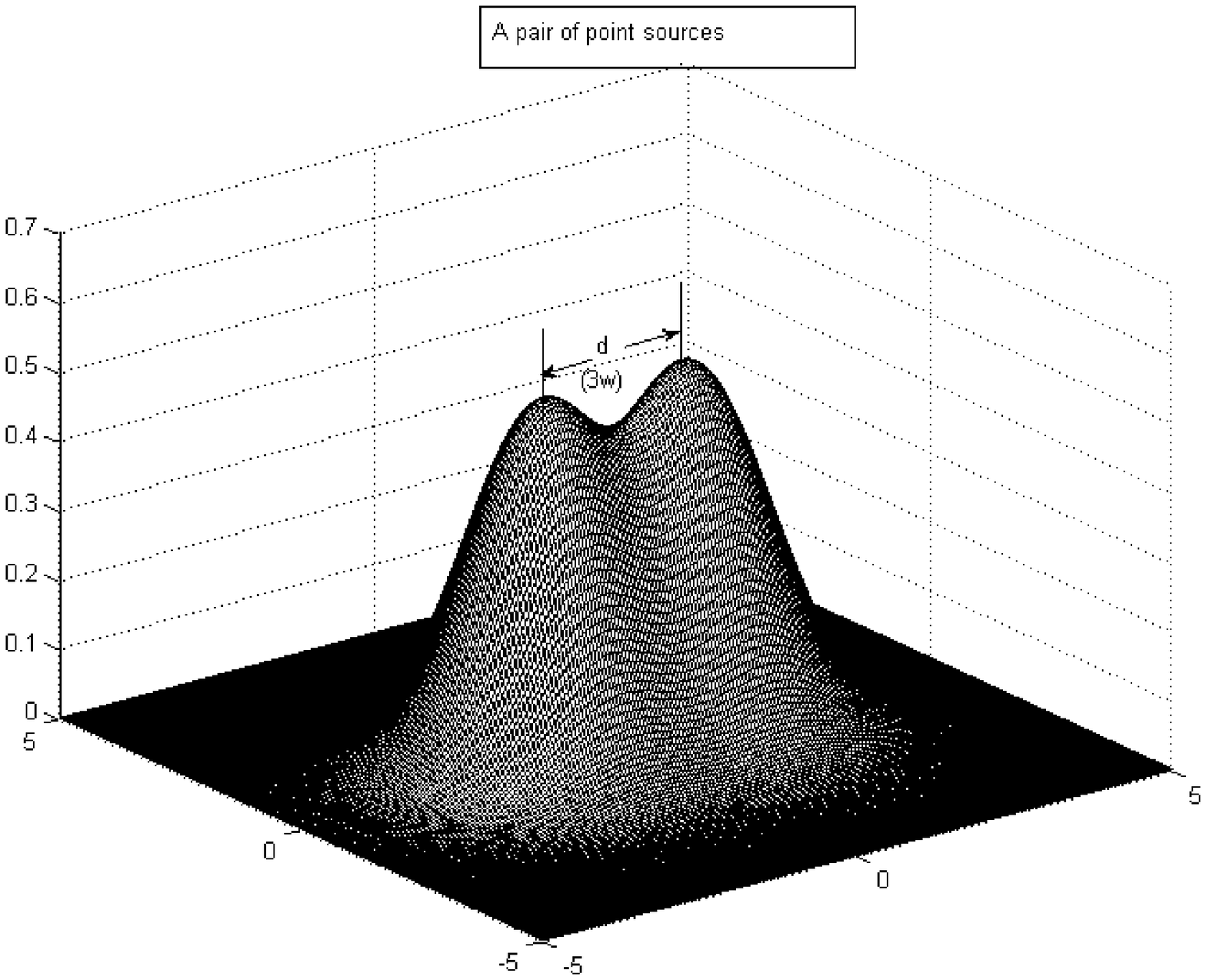}}
\caption{\label{fig:f1} The image of a pair of point sources of equal brightness
under the Gaussian-shaped PSF (\ref{e12}), for (a) $d/w=0.5$; (b) $d/w=2$; and (c) $d/w=3$.}
\end{figure}

With the elements of the data-mean matrices given by Eqs.~(\ref{e14}), 
let us now expand each term in the numerator $Q$ of the MPE expression, defined by relation (\ref{e9a}), in a power series in $K$, 
interchange the order of the resulting infinite sum and the pixel sum, and then perform the pixel sum 
by means of its continuous version. We arrive in this way at the following expression for $Q$:
\ba
\label{e15}
Q={K^2\Delta A\over (2\pi w^2)(\sigma^2+b)^{3/2}}&\sum_{n=0}^\infty {(-1)^n (2n+1)!! \over n!(n+2)}
\left[{K\,\Delta A\over 2(\sigma^2+b)2\pi w^2}\right]^n\nonumber\\
&\times\left[1+{1\over 2}e^{-\displaystyle{nd^2\over 4(n+2)w^2}}
- 2\,e^{-\displaystyle{(n+1)d^2\over 8(n+2)w^2}}
+{1\over 2}\,e^{-\displaystyle{d^2\over 4w^2}}\right],
\end{align}
in which the double factorial is defined as $(2n+1)!!= 1.3\ldots (2n+1)$, namely the product of all odd integers from 1 to $(2n+1)$.
The evaluation of simple Gaussian integrals needed to arrive at this result is presented in Appendix A. 
Expression (\ref{e15}) may be further simplified in the limiting case of interest, namely when $d<<w$,
by expanding the exponentials inside the square brackets in powers of $d/w$. The lowest-order term that survives in such an expansion
is quartic and, following straightforward algebra, shown to have the value,
$$\left({d^2\over 8w^2}\right)^2{(n^2+2n+3)\over (n+2)^2},$$  
so to this order the following expression for $Q$ results:
\ba
\label{e16}
Q={K^2\Delta A (d/w)^4\over (128\pi w^2)(\sigma^2+b)^{3/2}}&\sum_{n=0}^\infty {(-1)^n (2n+1)!! \over n!(n+2)}\nonumber\\
&\times {(n^2+2n+3)\over (n+2)^2}  \left[{K\,\Delta A\over 2(\sigma^2+b)2\pi w^2}\right]^n.
\end{align}
By means of the identity,
$${n^2+2n+3\over (n+2)^2} = 1-{2\over n+2}+{3\over (n+2)^2},$$
we may express $Q$ as 
\be
\label{e17}
Q={K^2\Delta A(d/w)^4\over (128\pi w^2)(\sigma^2+b)^{3/2}}[q_1(u)-2\,q_2(u)+3\,q_3(u)],
\ee
where the three $q$'s and $u$ are defined by the relation
\be
\label{e18}
q_k(u)\defeq \sum_{n=0}^\infty {(-1)^n (2n+1)!!\over 2^n n!(n+2)^k} u^n,\ \ \ u\defeq {K\Delta A\over 2\pi w^2 (\sigma^2+b)}.
\ee

The quantity $q_k$ may be evaluated recursively for increasing values of $k$
by a simple mathematical trick. Note first that $u^2q_k(u)$ is simply the integral, from 0 to $u$, of the sum
$$\sum_{n=0}^\infty {(-1)^n (2n+1)!!\over 2^n n!(n+2)^{k-1}} v^{n+1},$$
which is simply $v$ times $q_{k-1}(v)$,  
\be
\label{e19}
u^2 q_k(u) = \int_0^u v\, q_{k-1}(v) \, du.
\ee
Since $q_0(u)$ is simply the power-series expansion of $(1+u)^{-3/2}$, 
valid for $|u|<1$, 
we may evaluate the $q_k$ successively for increasing integer values of $k$ as
\ba
\label{e20}
u^2 q_1(u) &= \int_0^u dv {v\over (1+v)^{3/2}} \nonumber\\
           &=\int_0^u dv\left[{1\over (1+v)^{1/2}} -{1\over (1+v)^{3/2}}\right]\nonumber\\
           &=2(1+u)^{1/2}+2(1+u)^{-1/2} -4;
\end{align}
\ba
\label{e21}
u^2 q_2(u) &= 2\int_0^u dv\, v^{-1} [(1+v)^{1/2}+ (1+v)^{-1/2} -2]\nonumber\\
           &=4\left(\sqrt{1+u}-1\right) - 8 \ln \left({1+\sqrt{1+u}\over 2}\right);
\end{align}
and
\be
\label{e22}
u^2 q_3(u) = 4\int_0^u {dv \over v} \left[(1+v)^{1/2}-1\right]-8\int_0^u {dv\over v}\ln \left({1+\sqrt{1+v}\over 2}\right).
\ee
The details of the evaluation of integral (\ref{e21}) are presented in Appendix B, while integral (\ref{e22}), 
obtained by a substitution of expression (\ref{e21}) into relation (\ref{e22}) for $k=3$, 
must be evaluated numerically. 

These expressions for the various $q$'s when computed and substituted into expression (\ref{e17}) 
evaluate $Q$ fully. Note that while the power series expansions on which expression (\ref{e17}) is
based are valid only when $|u|<1$, the expressions (\ref{e20})-(\ref{e22}) are valid 
for arbitrary values of $u$ except where they fail to be analytic. 
This is guaranteed by the principle of analytic continuation \cite{CKP66}.

Expression (\ref{e9a}) for $R$, which determines the denominator of the MPE expression (\ref{e9}),
may be evaluated quite similarly as $Q$. In the small-spacing limit, $d/w<<1$, one
may again substitute $K\, H_{1ij}$ for $x_{1ij}$
and $K\, H_{2ij}$ for $x_{2ij}$, in that sum expression for $R$, then expand each term 
in the resulting expression in powers of $K$, interchange the power series sum with the 
pixel sum, and evaluate the latter, which in view of expressions (\ref{e14}) for $H_{1ij}$
and $H_{2ij}$ involves only Gaussian functions, approximately by converting it into an integral 
over the full $\xi\eta$ plane. Term by term, the Gaussian integrals are the same ones we evaluated
for $Q$, so their lowest-order limiting form in powers of $d/w$ is again quartic, and     
the following expression for $R$ results:
\ba
\label{e23}
R={K^2\Delta A (d/w)^4\over (128\pi w^2)(\sigma^2+b)^2}&\sum_{n=0}^\infty {(-1)^n (n+1) \over (n+2)}\nonumber\\
&\times {(n^2+2n+3)\over (n+2)^2}  \left[{K\,\Delta A\over 2\pi w^2 (\sigma^2+b)}\right]^n.
\end{align}
By means of the identity,
$${n^2+2n+3\over (n+2)^2} = 1-{2\over n+2}+{3\over (n+2)^2},$$
we may express $R$ as 
\be
\label{e24}
R={K^2\Delta A(d/w)^4\over (128\pi w^2)(\sigma^2+b)^2}[r_1(u)-2\,r_2(u)+3\,r_3(u)],
\ee
where $u$ is defined in Eq.~(20) and the three $r$'s are defined as the power series
\be
\label{e25}
r_k(u)\defeq \sum_{n=0}^\infty {(-1)^n (n+1)\over(n+2)^k} u^n, \ \ k=1,2,3.
\ee

Like the $q$'s, the $r$'s too may be evaluated recursively by means of an analogous integral relation they obey, namely
\be
\label{e26}
u^2 r_k(u) =\int_0^u v\, r_{k-1}(v)\, dv.
\ee
Since $r_0(u)$ given by (\ref{e25}) for $k=0$ is simply the Taylor expansion of $(1+u)^{-2}$, we have
\ba
\label{e27}
r_1(u) = &{1\over u^2}\int_0^u {v\over (1+v)^2} dv \nonumber\\
           = & {1\over u^2}\int_0^u \left[{1\over 1+v}-{1\over (1+v)^2}\right]dv\nonumber\\
           = & {1\over u^2}\left[\ln (1+u) +{1\over 1+u}-1\right] = {1\over u^2}\ln(1+u)-{1\over u(1+u)}.
\end{align}
From this expression for $r_1(u)$, we may now express $r_2(u)$ as the integral
\ba
\label{e28}
r_2(u) = &{1\over u^2}\int_0^u dv \left[{\ln (1+v)\over v}-{1\over (1+v)}\right]\nonumber\\
       = &{1\over u^2}\left[\int_0^u dv {\ln (1+v)\over v}-\ln (1+u)\right],
\end{align}
and, recursively, $r_3(u)$ as
\ba
\label{e29}
r_3(u) = &{1\over u^2}\int_0^u {dv\over v}\left[\int_0^v dw {\ln (1+w)\over w}-\ln (1+v)\right]\nonumber\\
       = & {1\over u^2}\int_0^u (\ln u-\ln v) {\ln (1+v)\over v} - {1\over u^2}\int_0^u {dv\over v}\ln(1+v),
\end{align}
where to reach the second equality we needed to perform an integration by parts of the double integral in the first equality.
These integral forms for $r_2$ and $r_3$ may be evaluated numerically, and $R$ given by 
Eq.~(\ref{e24}) thus fully calculated, with validity guaranteed 
for arbitrary values of the argument $u$ by analytic continuation. 

In terms of the quantities $q_k$ and $r_k$ we have just calculated, we may now express the argument of the 
erfc function in the MPE expression (\ref{e9}) via relations (\ref{e17}) and (\ref{e24}) as
\be
\label{e30}
{\norm{U_0} \over \sqrt{2}}={K\Delta A^{1/2}(d/w)^2\over 16[(4\pi w^2)(\sigma^2+b)]^{1/2}}
{[q_1(u)-2q_2(u)+3q_3(u)]\over [r_1(u)-2r_2(u)+3r_3(u)]^{1/2}},
\ee
in which $u$, defined by Eq.~(\ref{e18})
as the ratio of the characteristic number of mean signal photons per pixel, $K \Delta A/(2\pi w^2)$, and the
sum of the background and sensor noise variances per pixel, $\sigma^2 +b$, is a measure of the SNR for the problem.
We take $b>>\sigma^2$ in all our numerical evaluations, in which case $u$
may be interpreted as the signal-to-background ratio (SBR).  
We now develop limiting analytical forms for the right-hand side (RHS) of this expression
in the photon-signal-dominated and background-dominated regimes, given by $u>>1$ and
$u<<1$, respectively. 

\subsection{Photon-Signal-Dominated Regime, $u>>1$}

In this regime, we may use asymptotic forms of the various $p$'s and $q$'s occurring in 
expression (\ref{e30}). The following asymptotic forms, as we show in Appendix C, are obtained 
in the limit $u\to \infty$:
\ba
\label{e31}
&q_1(u) \sim {2\over u^{3/2}}; \ \  q_2(u) \sim {4\over u^{3/2}}; \ \ q_3(u) \sim  {8\over u^{3/2}};\nonumber\\
&r_1(u)\sim {\ln u\over u^2}; \ \ r_2(u) \sim {1\over 2u^2} [(\ln u)^2-2\ln u];\ \ r_3(u)\sim {1\over u^2}
\left[{(\ln u)^3\over 6}-{(\ln u)^2\over 2}\right].
\end{align}
With these asymptotic forms and definition (\ref{e18}) for $u$, expression (\ref{e30}) may be approximated as
\ba
\label{e32}
{\norm{U_0} \over \sqrt{2}}=&{K\Delta A^{1/2}(d/w)^2\over 16[(4\pi w^2)(\sigma^2+b)]^{1/2}}
{18\sqrt{2}\over \sqrt{u}\ln^{3/2} (u)}\nonumber\\
=& {9\over 8}{\sqrt{K} d^2/w^2\over \ln^{3/2}(u)}.
\end{align}

Let us set the threshold on the MPE at $p$ as the minimum requirement on the fidelity of Bayesian discrimination between
the single-point-source and symmetric binary-source hypotheses. A typical value taken for $p$ is 0.05, corresponding to 
a statistical CL of 95\%, but one can adjust $p$ according to the stringency of the application. Equating the RHS of relation (\ref{e10})
to $p$ and solving for the argument of the erfc function in terms of the inverse function, which we may
denote erfc$^{-1}$, we require that $\norm{U_0}/\sqrt{2}$ have the minimum value
\be
\label{e33}
{\min\norm{U_0}\over\sqrt{2}} = \erfc^{-1}(2p)
\ee
and thus, from the asymptotically valid relation (\ref{e32}), arrive at the following implicit value of the 
minimum photon number, $K_{min}$, needed to achieve this fidelity:
\be
\label{e34}
K_{min}= 
\left[{8\, \erfc^{-1}(2p)\over 9}\right]^2{w^4\over d^4}\ln^3\left[{K_{min}\Delta A/(2\pi w^2)\over (b+\sigma^2)}\right].
\ee

Expression (\ref{e34}) exhibits a nearly quartic scaling of the minimum photon number needed to discriminate
a point-source pair from its single-point-source equivalent, as a function of the inverse spacing $d^{-1}$ 
between the source pair. Specifically, this scaling is given by the ratio $(w/d)^4$, namely
the fourth power of the ratio of the characteristic width of the PSF and the spacing between the source
pair being resolved, that is modified by a logarithmic dependence on the SNR. The latter factor
increases relatively slowly with increasing $K_{min}$, but tends to exacerbate somewhat a purely quartic scaling of $K_{min}$
as higher and higher values of the pair-OSR factor, $w/d$, are sought.    
Such logarithmic factors have not been predicted by previous researchers \cite{Helstrom69,Lucy92,SM04,Smith05},
as their analyses have not been sufficiently comprehensive in treating the full range of combined noise 
statistics. Further, the analyses of Ref.~\cite{Helstrom69,SM04} assumed essentially white 
Gaussian additive noise, rather than photon-number-dominated Poisson noise treated here, and their
power SNR scales quadratically, rather than linearly, with the photon number. This difference of the
noise statistics accounts for their quadratic, rather than our essentially quartic, scaling of $K_{min}$
on the degree of pair OSR sought. 

In spite of its modification by a logarithmic correction, the quartic scaling of $K_{min}$ 
is considerably more modest than the eighth-power scaling predicted
by Lucy \cite{Lucy92} for a similar problem. As we have argued earlier, the assumption of
a single point source in our analysis, rather than a single equivalent but extended source in Lucy's analysis,
for the ``null" hypothesis of an unresolved source potentially provides more constraining information that
an algorithm can make essential use of to resolve the source pair at a more modest signal strength.    

\subsection{Background-Noise-Dominated Regime, $u<<1$}

In the limit $u\to 0$, the various $q$ and $r$ functions tend to the same values, $q_k=r_k=(1/2)^k$,
so the ratio involving them in expression (\ref{e30}) simplifies to $(3/8)^{1/2}$
\be
\label{e35}
{\norm{U_0} \over \sqrt{2}}=\sqrt{3\over 8}{K\Delta A^{1/2}(d/w)^2\over 16[(4\pi w^2)(\sigma^2+b)]^{1/2}},
\ee
so the minimum photon number needed for a pair superresolution factor of amount $w/d$ is now expected to scale
only quadratically with that factor, without any logarithmic corrections. For a fidelity denoted by the MPE threshold $p$, we
may perform a similar analysis as for the photon-signal-dominated regime of the previous sub-section
to arrive at the following expression for $K_{min}$:
\be
\label{e36}
{K_{min}\Delta A^{1/2}(d/w)^2\over 16[(4\pi w^2)(\sigma^2+b)]^{1/2}},
=\sqrt{8\over 3}\,\erfc^{-1}(2p).
\ee

\section{Numerical Results and Discussion}

In Figs.~2(a)-(c), we plot the RHS of the exact result (\ref{e30}) as a function of the photon number 
$K$ for three different background variance levels, $b$, and 
illustrate the transition from the quadratic scaling of the background-dominated regime
to the approximately quartic scaling of the signal-dominated regime of operation of our Gaussian-PSF-based imager.
In each plot, the sensor noise variance $\sigma^2$ has been set equal to 1, while the ratio
of the characteristic area under the PSF to the pixel area, $2\pi w^2/\Delta A$, is set to 100.
\begin{figure}
\centering
\subfloat[]
{\includegraphics[width=3in]{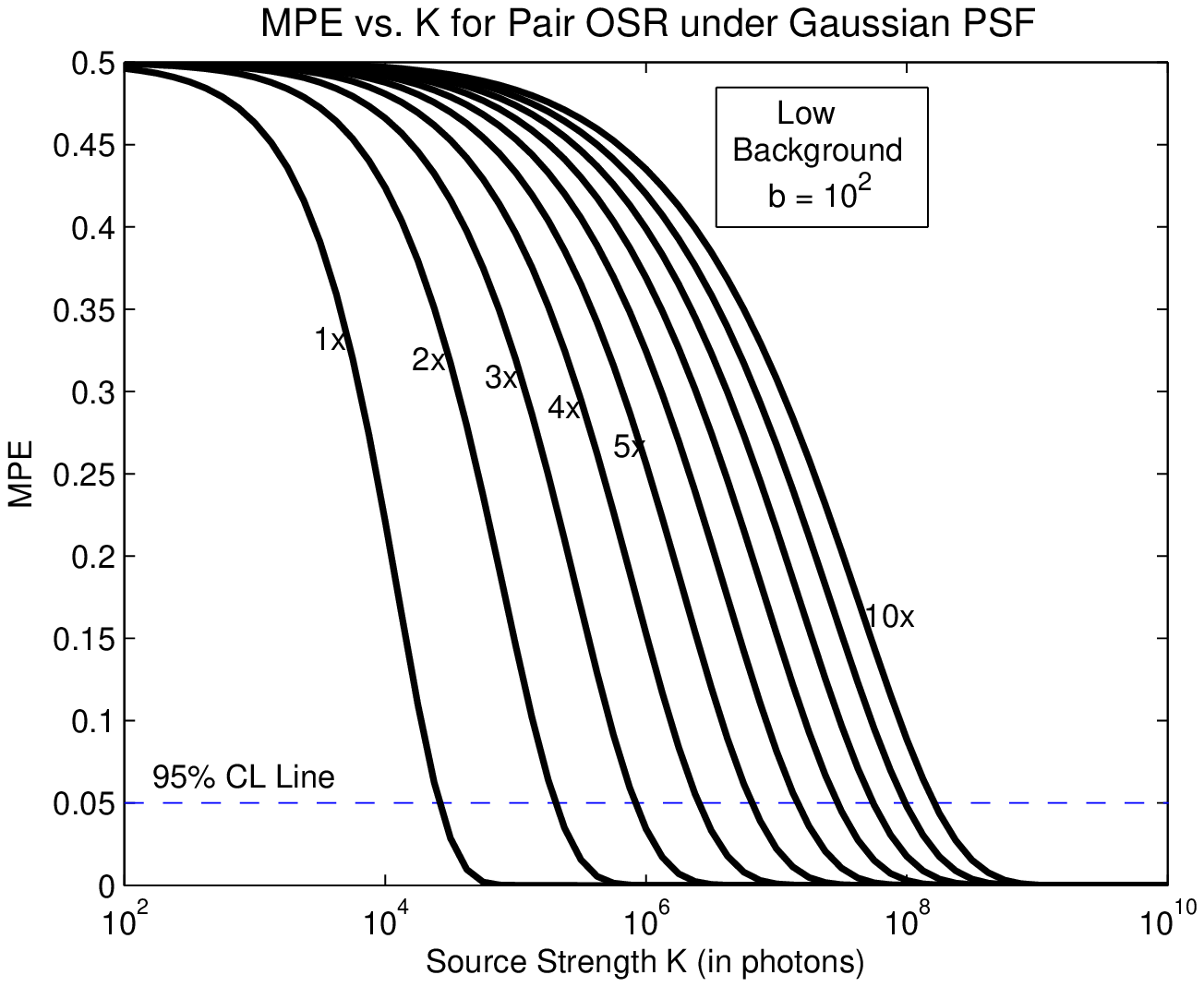}}
\subfloat[]
{\includegraphics[width=3in]{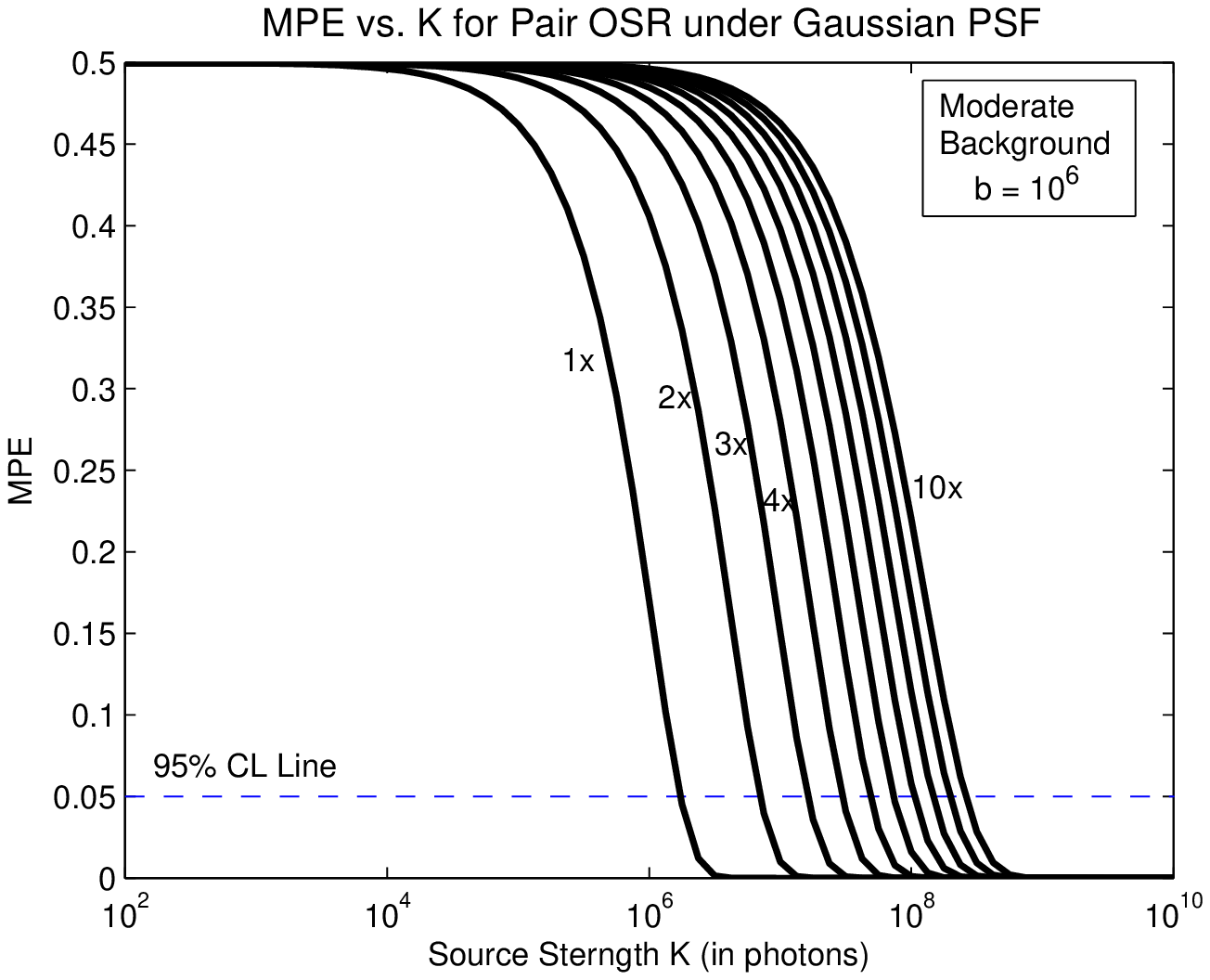}}
\hspace{.2cm}
\subfloat[]
{\includegraphics[width=3in]{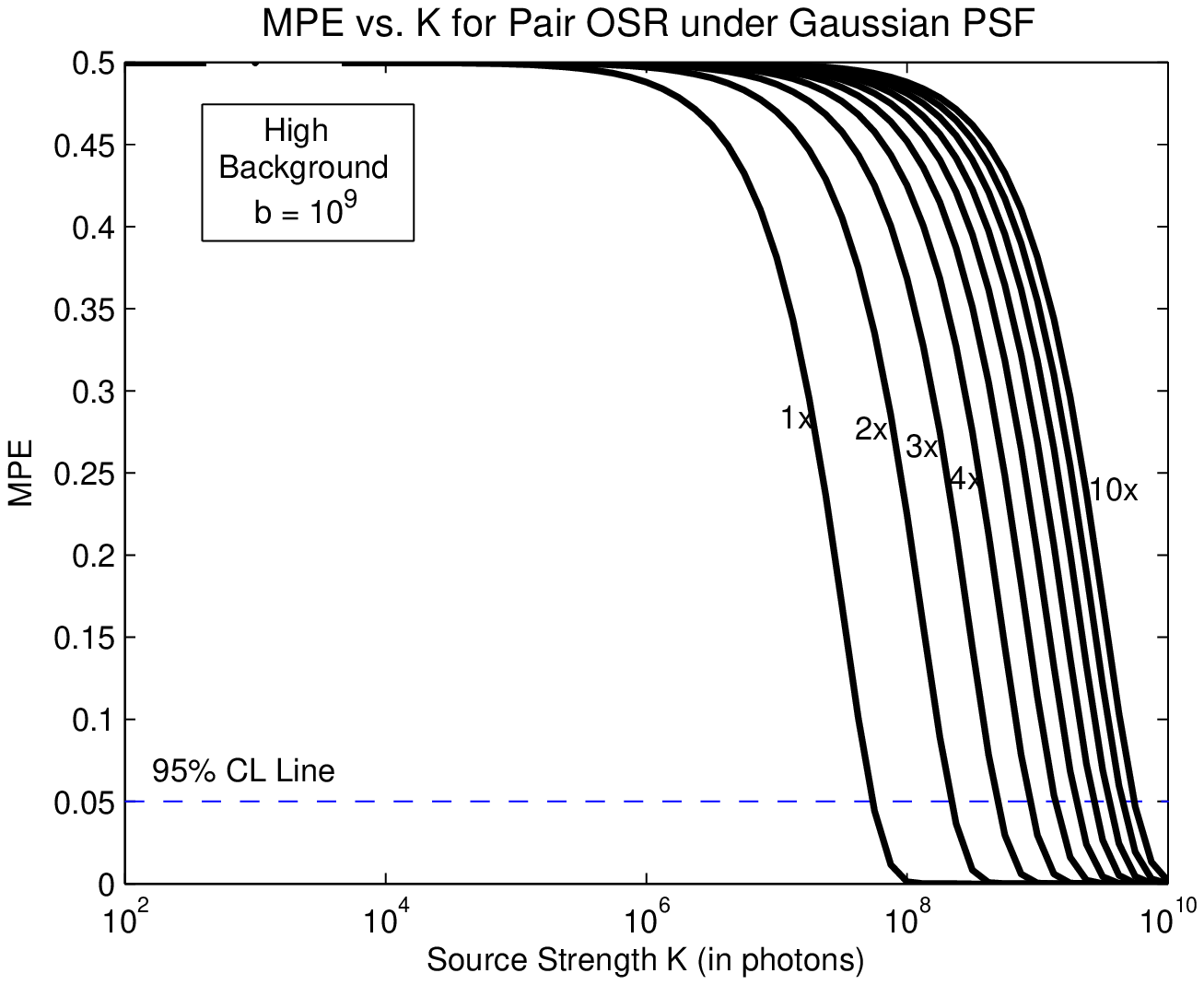}}
\caption{\label{fig:f2} A plot of MPE vs. signal strength for three different background levels. (a) $b=10^2$; (b) $b=10^6$; and
(c) $b=10^9$.}
\end{figure}
For the case of a low background level, $b=10^2$, even at the lowest source strength plotted in Fig.~2(a), namely $K=10^2$,
the signal-to-background photon ratio (SBR) $u$, defined by relation (\ref{e18}), is at its lowest value comparable to 1. At much
higher source strengths of interest, $K>10^4$, $u>>1$ and the plot displays
the decrease of the MPE with increasing signal strength appropriate to the signal-dominated regime in which the
approximation (\ref{e32}) for the argument of the erfc function is quite accurate. 
The different curves in the plots refer to different
values of the OSR ratio $w/d$, starting at 2 (denoted as 1x) and increasing to 20 (denoted as 10x),
with the higher $w/d$ values requiring higher source strengths to bring the error down to the CL threshold. 
The bottom figure, Fig.~2(c), displays the MPE behavior in the opposite limit of background-dominated operation in which
at the largest source strength, $K=10^{11}$, the ratio $u$ is 1, and for all others it is less than 1, being in fact considerably
smaller than 1 over the range of source strengths over which the MPE decreases from its highest value of 0.5 toward the 
statistical-confidence-level (CL) threshold of 0.05. 
The upper right figure, Fig.~2(b), displays the results for the
case for which its left half represents the background-dominated regime of operation
and the right half the signal-dominated regime. Note an appreciable but expected rightward shift in the MPE curves as
the spatially uniform background level increases from one figure to the next. 
As the background level increases, it becomes increasingly harder, requiring increasingly higher source
strengths, to discriminate between the null and alternative hypotheses. 
    
The variation of the minimum photon number required to achieve the CL threshold as a function of the OSR ratio,
$w/d$, is plotted in Fig.~3 for the three different background levels of Fig.~2. The values of this number were simply read
off from the intersections of the MPE curves in Figs.~2 with the CL threshold line. The doubly logarithmic plots
show the expected nearly quartic and quadratic scaling with $w/d$ for the photon-dominated and background-dominated
regimes, while for the intermediate background level, $b=10^6$, the scaling is nearly quadratic for the smaller $d/w$ values
for which $K_{min}$ is smaller than or comparable to $(2\pi w^2/\Delta A)$, or 100, times $b$, but for larger $w/d$,
the slow change of slope consistent with the nearly quartic scaling of the signal-dominated regime can be discerned. 
The transition from slope-2 to slope-4 is rather gradual, taking several decades of increase in $K_{min}$ to complete.
\begin{figure}
\centering
{\includegraphics[width=4in]{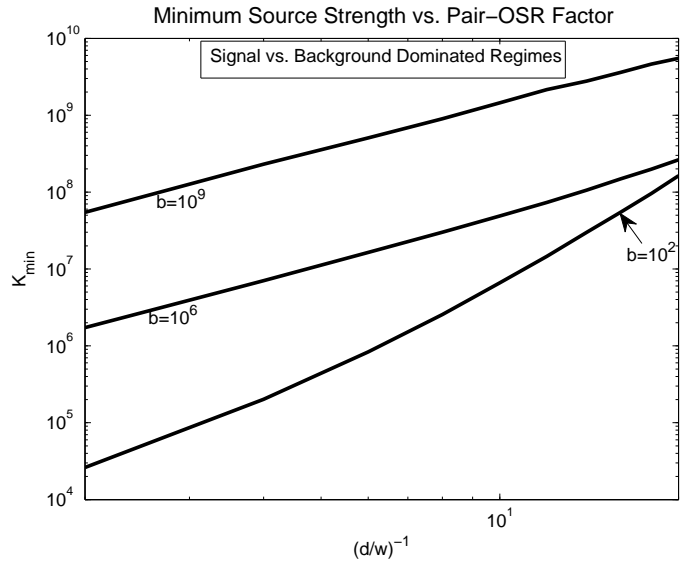}}
\caption{\label{fig:f3} Log-log plots of the minimum source strength, $K_{min}$, vs. the OSR ratio, $w/d$,
for the same three different background levels as in Figs.~2.}
\end{figure}

It is rather remarkable that one can achieve {\it any} OSR at all when the spatially uniform mean background level
and its shot-noise fluctuations dominate, as in Fig.~2(c), 
the signal part of the data with its information-bearing
spatial variations that are consistent with the background-free image of a separated source pair.
The higher source strengths needed when even relatively small OSR ratios are sought, as seen in Fig.~3, confirm the difficulty of
attaining OSR in the background-dominated regime.
The change of slope of the $\ln w/d$ vs. $\ln K_{min}$ curve from 4 to 2 as the background gets stronger
relative to the signal must mean that the various curves in Fig.~3 must asymptote to the same limit as the 
degree of sought OSR and the corresponding $K_{min}$ are increased to still greater values not shown in the plot. 

It is worth noting that the sensor and background variances per pixel occur together in all our expressions
as $\sigma^2+b$. In the pseudo-Gaussian approximation for the PDF for combined sensor, background, and signal photon-number
fluctuations, the various noise sources are equivalent. However, that is not so in a more accurate treatment needed to incorporate 
either background or signal photon-number fluctuations at much lower strengths of order 10 or lower, where their discrete 
Poisson statistics can no longer be approximated well by a continuous Gaussian PDF.

\section{Conclusions}

In this paper we have presented a detailed Bayesian error analysis based on binary-hypothesis testing (BHT) to 
derive expressions for the minimum combined source strength needed to superresolve a pair of closely spaced
point sources located in the plane of best focus from their image formed under a Gaussian PSF approximation. 
The statistical metric we have used here to determine this minimum value of the source strength 
needed for pair OSR is the minimum probability of error (MPE) on a successful BHT protocol. Specifically, we considered 
the pair to be resolved if the MPE for the associated BHT problem falls below a small threshold value, taken
here to be 0.05. Our calculations are done for a variety of operating conditions characterized by arbitrary 
values of the source signal strength, background photon count, and sensor noise variance.

Our detailed quantitative calculations predict an approximately quartic dependence of the 
minimum source strength on the reciprocal of the pair spacing in the regime where the
signal dominates the background. This scaling of source strength with respect to the inverse spacing 
is in fact slightly steeper than quartic via a multiplicative correction that is logarithmic in
the ratio of the signal and background levels.
In the opposite limit of the background-dominated regime of operation, this scaling
is more modestly quadratic. 

The problem of pair OSR when the sources are along the optical axis and thus in a common line of sight (LOS)
is expected to entail much steeper scaling of minimum source power with inverse spacing in both the 
signal-dominated and background-dominated regimes. The main reason for this difference of on-axis, or longitudinal,
OSR from the transverse OSR treated in the present paper is that in the former case, unlike the latter,  
the PSF has no first order sensitivity on the spacing, which must imply more stringent 
requirements on any LOS pair OSR. However, the mathematical divergences seen in local first-order analyses using 
the Cram\'er-Rao lower bound on unbiased estimation are expected to be moderated in a Bayesian analysis
that includes in effect non-vanishing higher-order sensitivities of the data PDF on the parameter being estimated, here 
the pair spacing. This problem will be treated in a subsequent paper. 

\section*{Acknowledgments}
Helpful conversations with Rakesh Kumar, Srikanth Narravula, Julian Antolin, Henry Pang, and Keith Lidke 
are gratefully acknowledged. The work reported here was supported in part by AFOSR under grant numbers
FA9550-09-1-0495 and FA9550-11-1-0194.

\appendix
\section{Certain Gaussian Integrals}
The following integral over the full $\xi\eta$ plane must be evaluated to arrive at expression (\ref{e15}):
\ba
\label{a1}
\int d\xi d\eta &\exp\left[-{n(\xi^2+\eta^2)\over 2w^2}\right] \left\{\exp\left[-{\xi^2+\eta^2\over 2w^2}\right]\right.\nonumber\\
                &\left. -{1\over 2}\exp\left[-{(\xi-d/2)^2+\eta^2\over 2w^2}\right]
                -{1\over 2}\exp\left[-{(\xi+d/2)^2+\eta^2\over 2w^2}\right]\right\}^2.
\end{align}
Squaring the terms inside the braces and then multiplying out the Gaussian functions in the integrand
yields four different kinds of Gaussian integrals, each over the $\xi\eta$ plane, namely
\ba
\label{a2}
I_0\defeq &\int \exp\left[-{(\xi^2+\eta^2)(n+2)\over 2w^2}\right] d\xi d\eta;\nonumber\\
I_\pm\defeq &\int \exp\left[-{(n+1)\xi^2+(\xi\mp d/2)^2+(n+2)\eta^2\over 2w^2}\right] d\xi d\eta;\nonumber\\
J_\pm\defeq &\int \exp\left[-{n\xi^2+2(\xi\mp d/2)^2+(n+2)\eta^2\over 2w^2}\right] d\xi d\eta; \ {\rm and}\nonumber\\
J_0\defeq &\int \exp\left[-{n\xi^2+(\xi- d/2)^2+(\xi+d/2)^2+(n+2)\eta^2\over 2w^2}\right] d\xi d\eta.
\end{align}

The first of these integrals is simply evaluated as
\be
\label{a3}
I_0 ={2\pi w^2\over n+2}.
\ee
The remaining integrals are evaluated by ``completing the square" in each Gaussian exponent that contains unshifted and
shifted quadratic expressions. Thus, for example,
\be
\label{a4}
(n+1)\xi^2+(\xi\mp d/2)^2= (n+2)\left[\xi\mp {d\over 2(n+2)}\right]^2+{(n+1)d^2\over 4(n+2)},
\ee
a trick that, when used in the exponent of the second of the Gaussian integrals, $I_\pm$, in relation (\ref{a2}),
followed by an appropriate finite shift of the infinite range of the integral, enables
us to evaluate the $\xi$ part of the integral. The $\eta$ part of the double integral in each of these 
integral expressions is always the same, and evaluates to $\sqrt{2\pi w^2/(n+2)}$. Accounting for the 
left-over terms like $(n+1)d^2/[4(n+2)]$ in each exponent of the $\xi$-dependent integrand
then yields the following evaluations for the remaining integrals: 
\ba
\label{a5}
I_\pm=&{2\pi w^2\over (n+2)}\exp\left[-{d^2\over 8w^2}\left({n+1\over n+2}\right)\right];\nonumber\\
J_\pm=&{2\pi w^2\over (n+2)}\exp\left[-{d^2\over 4w^2}\left({n\over n+2}\right)\right]; \ {\rm and}\nonumber\\
J_0=&{2\pi w^2\over (n+2)}\exp\left[-{d^2\over 4w^2}\right].
\end{align}

\section{Evaluation of Integral (\ref{e21})}

By writing, $v=\sinh^2\alpha$, and noting that $1+v=\cosh^2\alpha$, we may simplify integral (\ref{e21}) as
\ba
\label{b1}
2&\int_0^{\sinh^{-1}\sqrt{u}}d\alpha{(\cosh\alpha-1)^2\over \sinh\alpha} \nonumber\\
=&4\int_0^{\sinh^{-1}\sqrt{u}}d\alpha{\sinh^3(\alpha/2)\over \cosh(\alpha/2)}, 
\end{align}
where we obtain the second line from the first by means of the identities, $\cosh\alpha=1+2\sinh^2(\alpha/2)$ and 
$\sinh\alpha=2\sinh(\alpha/2)\cosh(\alpha/2)$. We now make another substitution, $\beta=\cosh(\alpha/2)$, and note
that $d\beta=\sinh(\alpha/2)d\alpha/2$ to reduce the above integral expression to the form,
\be
\label{b2}
8\int_1^{\cosh(\sinh^{-1}\sqrt{u}/2)} d\beta {\beta^2-1\over\beta}.
\ee
This integral is easily evaluated as 
\ba
\label{b3}
4&\left[\cosh^2(\sinh^{-1}\sqrt{u}/2)-1\right] -8\ln\cosh(\sinh^{-1}\sqrt{u}/2)\nonumber\\
=&4\sinh^2(\sinh^{-1}\sqrt{u}/2)-4\ln\left[{1+\cosh(\sinh^{-1}\sqrt{u})\over 2}\right]\nonumber\\
=2&\left[\cosh(\sinh^{-1}\sqrt{u})-1\right]-4\ln\left({1+\sqrt{1+u}\over 2}\right)\nonumber\\
=2&\left[\sqrt{1+u}-1\right]-4\ln\left({1+\sqrt{1+u}\over 2}\right),
\end{align}
where we used simple hyperbolic-function identities, $\cosh^2\alpha=1+\sinh^2\alpha$ and
$1+2\sinh^2(\alpha/2)=2\cosh^2(\alpha/2)-1=\cosh\alpha$, to 
reduce the various expressions to their final form. 

\section{Asymptotic Forms of the Various $q$ and $r$ Functions}

From expressions (\ref{e20}) and (\ref{e21}), the large-$u$ approximations for $q_1(u)$ and $q_2(u)$
follow quite simply, since for $u>>1$ any positive power of $u$ dominates any constants or logarithms of $u$ or
negative powers of $u$,
\be
\label{c1}
q_1(u>>1)\sim {2\over u^{3/2}},\ \ q_2(u>>1)\sim {4\over u^{3/2}}.
\ee
For large $u$, the integrals in Eq.~(\ref{e22}) representing $u^2q_3(u)$ may be approximated by 
approximating their integrands near the upper limit $u$,
\be
\label{c2}
{[(1+v)^{1/2}-1]\over v}\sim v^{-1/2},\ \ \ln \left({1+\sqrt{1+v}\over 2}\right)\sim {1\over 2}\ln v,
\ee
so the two integrals may be evaluated straightforwardly near the upper limit and the following expression
for $q_3(u)$, valid for $u>>1$, results:
\be
\label{c3}
q_3(u>>1) \sim {8\over u^{3/2}} - {2\over u^2}\ln^2(u)\ \sim {8\over u^{3/2}},
\ee
where we have ignored, in the second approximate equality, the second term of the first
approximate equality as being logarithmically smaller for sufficiently large $u$ for which $u^{-1/2}ln^2 u << 4$.   

Similar considerations give us the asymptotic forms for $r_k(u)$, $k=1,2,3$. From the last equality in Eq.~(\ref{e27}),
it follows that $r_1(u>>1)\sim u^{-2}\ln u.$ We may now determine the
approximate form for $r_2(u)$ for $u>>1$ by approximating the integrand of the integral in expresssion (\ref{e28})
as $v^{-1}\ln v$, whose integral is easily evaluated near the upper limit as $(1/2)\ln^2 u$. The following asymptotically valid
form for $r_2(u)$ is thus obtained:
\be
\label{c4}
r_2(u>>1)\sim {1\over 2u^2}(\ln^2 u -2\ln u).
\ee
Finally, a similar approximation, $\ln (1+v)\approx \ln v$, in the integrand of expression (\ref{e29}) for $r_3(u)$
enables us to evaluate the integral near its upper limit $u$, for $u>>1$, as
\ba
\label{c5}
r_3(u)&\sim {1\over u^2} \left({\ln^3u\over 2}-{\ln^3u\over 3}-{\ln^2u\over 2}\right)\nonumber\\
&= {1\over u^2} \left({\ln^3u\over 6}-{\ln^2u\over 2}\right).
\end{align}
In deriving expressions (\ref{c4}) and (\ref{c5}),
we have been careful to retain terms that are logarithmic in $u$ as being comparable to numerical constants of order 1.

\end{document}